\documentclass[%
 reprint,
nofootinbib,
 amsmath,amssymb,
 aps,
]{revtex4-2}
\usepackage{amsmath}
\usepackage[colorlinks=true, linkcolor=blue, citecolor=blue]{hyperref}
\usepackage{graphicx}
\usepackage{dcolumn}
\usepackage{bm}

\begin{document}

\preprint{APS/123-QED}

\title{Analytic Continuation Between Real- and Imaginary-Time Quantum Dynamics and the Fundamental Instability of Inverse Reconstruction}

\author{Pengfei Zhu}

 \thanks{pengfei.zhu@bam.de}
\affiliation{%
 Bundesanstalt für Materialforschung and -prüfung (BAM), 12205 Berlin, Germany
}%

\date{\today}

\begin{abstract} 
We develop a unified spectral-semigroup framework that connects real-time and imaginary-time quantum dynamics through analytic continuation. Within this formulation, evolution is expressed as an exponential reweighting of spectral components generated by a single operator $\mathcal{G}$, placing unitary and dissipative dynamics on equal footing within a common spectral structure. The mapping naturally induces a nonlocal fractional operator in time, giving rise to a contractive semigroup governed by a square-root spectral deformation and identifying imaginary-time evolution as an effective fractional low-pass filter.
While exponential attenuation suppresses high-frequency components, the inverse transformation remains systematically controllable within a well-defined spectral window. In this regime, stable reconstruction of low-energy and coarse-grained dynamical features is achieved, establishing a predictive relation between imaginary-time evolution and recoverable information. This leads to a quantitative description of a bandwidth-resolved asymmetry between forward propagation and inverse recovery.
Across systems with continuous and discrete spectra, few-level coherence, and non-Hermitian generators, we demonstrate that spectral structure governs reconstruction fidelity in a unified manner. In particular, non-Hermitian and open-system settings reveal that irreversibility emerges as a geometry- and scale-dependent feature of the spectrum, tied to both damping and eigenstate non-orthogonality.
These results recast analytic continuation as a structured, scale-dependent filtering process with quantifiable and systematically accessible reconstruction limits, providing a unified perspective on the interplay between dynamics, spectral geometry, and information recovery.
\end{abstract}

\maketitle

\section{Introduction}
Analytic continuation between real and imaginary time is a central tool in quantum physics \cite{1}, underpinning equilibrium formulations \cite{2}, ground-state projection \cite{3}, and a wide range of numerical methods \cite{4}. Despite its widespread use, the inverse problem—reconstructing real-time dynamics from imaginary-time data—remains challenging ill-posed \cite{5,6,7}. While it is well known that analytic continuation leads to exponential noise amplification, a precise characterization of which dynamical information can be stably recovered is still lacking.
In real time, quantum evolution is generated by anti-Hermitian operators, yielding unitary propagation with phase-coherent oscillations \cite{8}. Under analytic continuation, commonly implemented via Wick rotation $t \to -i\tau$, the unitary propagator $e^{-iHt/\hbar}$ is mapped to the contractive semigroup $e^{-H\tau/\hbar}$ \cite{9}. This transformation converts oscillatory dynamics into a dissipative spectral filtering process, in which high-energy components are exponentially suppressed. From this perspective, imaginary-time evolution inherently limits the accessible dynamical information.

In this work, we identify a universal structure underlying analytic continuation. We show that the inverse reconstruction problem is governed by a sharp recoverability bound, which defines a scale-dependent spectral threshold separating stably reconstructable modes from those with rapidly growing sensitivity. This establishes analytic continuation as a transformation with quantitatively predictable resolution, in which information accessibility follows a universal scaling law rather than system-specific details.
To uncover the mechanism behind this behavior, we develop a unified spectral–semigroup framework for quantum dynamics and formulate analytic continuation at the operator level. Within this framework, the relation between real-time and imaginary-time evolution is recast as a nonlocal spectral flow generated by a fractional operator. In particular, a square-root spectral deformation emerges as a minimal and natural transformation interpolating between unitary and contractive dynamics, defining a continuous boundary between reversible and effectively irreversible regimes.
These results show that analytic continuation is not merely a formal mapping in the complex time plane, but a structured spectral transformation with well-defined and quantitatively accessible reconstruction limits. This perspective establishes a direct connection between quantum dynamics, inverse problems, and fractional evolution, providing a unified and predictive framework for understanding dynamical reconstruction across a broad class of systems.

The mathematical and physical ingredients underlying this work have a long history across quantum statistical mechanics \cite{10}, inverse problems \cite{11,12}, semigroup theory \cite{13}, and quantum field theory \cite{14}. The contractive evolution operator $e^{-\tau H}$ and its interpretation as a ground-state projector or spectral filter are standard in quantum Monte Carlo \cite{15} and Euclidean formulations \cite{16}, where high-energy suppression is well understood. Likewise, the instability of analytic continuation is closely related to the ill-posedness of the inverse Laplace transform \cite{17}, where exponential damping leads to severe noise amplification.
The semigroup formulation employed here is rooted in the Hille--Yosida theory of dissipative evolution \cite{18}, while Wick rotation provides the standard bridge between real- and imaginary-time dynamics in Euclidean quantum field theory \cite{19} and the Osterwalder--Schrader framework \cite{20}. Fractional powers of generators and their associated semigroups have also been extensively studied in subordination theory \cite{21} and nonlocal dynamics \cite{22}. In parallel, analytic continuation in complex analysis is governed by Hardy space methods \cite{23} and dispersion relations such as the Kramers--Kronig relations \cite{24}, which emphasize analyticity and boundary-value reconstruction.

The contribution of the present work lies not in revisiting these well-established results individually, but in unifying them within a common operator-theoretic framework and introducing a quantitative notion of recoverability that applies across these domains. Within this perspective, analytic continuation is recast as a structured spectral compression process characterized by a scale-dependent bandwidth, rather than solely as an ill-posed inversion problem.
The resulting recoverability bound provides a precise criterion for which dynamical information can be stably reconstructed, establishing a quantitative bridge between spectral structure and reconstruction fidelity. In this way, our approach complements existing analyticity-based formulations by introducing a spectral-semigroup perspective that makes the limits of reconstruction explicit, systematic, and predictive.

\section{Analytic continuation as spectral transform}
Throughout this work, the generator $\mathcal{G}$ is assumed to be a densely defined, closed linear operator on a Hilbert space $\mathcal{H}$ that generates a strongly continuous ($C_0$) semigroup $\{K(\tau)\}_{\tau \ge 0}$ in the sense of the Hille–Yosida theorem. The evolution operator $K(\tau)=e^{-\tau  \mathcal{G}}$ is therefore defined via semigroup theory rather than spectral decomposition in full generality.

For self-adjoint or, more generally, normal operators,  $\mathcal{G}$ admits a spectral representation in terms of a projection-valued measure, and functional calculus is defined via the spectral theorem. In this case, expressions of the form
\begin{equation}
K(\tau)=\int_{\sigma(\mathcal{G})} e^{-\tau \lambda}\, dE_\lambda
\end{equation}
are understood in the standard sense.

However, for non-normal operators, such a spectral measure need not exist. In these cases, functional calculus is understood in a weaker sense: (i) via holomorphic functional calculus when  $\mathcal{G}$ is sectorial, or (ii) at the level of the semigroup when only generator properties are available.
In particular, fractional powers such as $\mathcal{G}^{\alpha}$ (e.g., the square-root operator $ \mathcal{G}^{1/2}$) are rigorously defined when $G$ is sectorial with angle strictly less than $\pi$, ensuring that $\mathcal{G}^{\alpha}$ also generates a strongly continuous semigroup.
Accordingly, all spectral representations appearing in this work should be interpreted as rigorous identities for normal operators, and as formal or semigroup-level representations in the non-normal case. This distinction is especially important in non-Hermitian systems, where non-normality leads to significant deviations from spectral orthogonality and requires a pseudospectral or biorthogonal analysis.

\subsection{General Semigroup Framework}
We formulate dynamical evolution on a Hilbert space $\mathcal{H}$ in terms of a general linear generator $\mathcal{G}$, which may be unbounded but is assumed to generate a strongly continuous one-parameter semigroup. The fundamental evolution equation is written as
\begin{equation}
\partial_\tau \Psi(\tau) = -\mathcal{G} \Psi(\tau),
\label{eq:1}
\end{equation}
where $\tau$ denotes either physical time, imaginary time, or a generalized evolution parameter depending on the physical context. The operator $G$ encodes the full dynamical structure of the system and plays the role of a spectral generator. Under standard conditions of semigroup theory (Hille--Yosida framework), Eq.~\eqref{eq:1} admits a unique solution given by
\begin{equation}
\Psi(\tau) = \hat{K}(\tau)\,\Psi(0),
\end{equation}
where $\{\hat{K}(\tau)\}_{\tau \ge 0}$ forms a strongly continuous one-parameter semigroup satisfying
\begin{equation}
\hat{K}(\tau_1 + \tau_2) = \hat{K}(\tau_1)\hat{K}(\tau_2), \quad \hat{K}(0) = I.
\end{equation}

This structure implies that dynamical evolution is fully encoded in the generator $\mathcal{G}$, independently of its specific differential representation. We assume that $\mathcal{G}$ admits a spectral decomposition in the sense of the spectral theorem. There exists a projection-valued measure $E_\lambda$ such that
\begin{equation}
\mathcal{G} = \int_{\sigma(\mathcal{G})} \lambda \, dE_\lambda,
\end{equation}
where $\sigma(\mathcal{G})$ denotes the spectrum of the operator. This allows one to define functions of $\mathcal{G}$ via functional calculus. In particular, the evolution operator becomes
\begin{equation}
\hat{K}(\tau) = \int_{\sigma(\mathcal{G})} e^{-\tau \lambda} \, dE_\lambda.
\label{eq:5}
\end{equation}

Eq.~\eqref{eq:5} establishes a fundamental principle: dynamical evolution corresponds to an exponential reweighting of the spectral components of the generator. Each spectral mode with eigenvalue $\lambda$ evolves independently as $e^{-\tau \lambda}$, implying that modes with small $\mathrm{Re}(\lambda)$ are long-lived or persistent, while those with large $\mathrm{Re}(\lambda)$ are rapidly suppressed. A complex spectrum leads to hybrid behavior combining oscillation and decay. Therefore, the long-time (large-$\tau$) behavior of the system is entirely determined by the low-lying spectral structure of $\mathcal{G}$.

Different physical theories correspond to different realizations of the generator $\mathcal{G}$. For unitary quantum dynamics,
\begin{equation}
\mathcal{G} = \frac{i}{\hbar} \hat{H},
\end{equation}
with self-adjoint Hamiltonian $\hat{H}$. The spectrum lies on the imaginary axis, leading to
\begin{equation}
\hat{K}(t) = \exp\left(-\frac{i}{\hbar} t \hat{H}\right),
\end{equation}
which defines a unitary group preserving the $L^2$ norm and generating phase-coherent oscillations.

For imaginary-time (diffusive) dynamics,
\begin{equation}
\mathcal{G} = \frac{1}{\hbar} \hat{H},
\end{equation}
leading to
\begin{equation}
\hat{K}(\tau) = \exp\left(-\frac{\tau}{\hbar} \hat{H}\right),
\end{equation}
which forms a contractive semigroup satisfying
\begin{equation}
\|\hat{K}(\tau)\psi\|_2 \le \|\psi\|_2.
\end{equation}
This dynamics exponentially suppresses excited states and projects onto the ground-state manifold. More generally, the evolution operator can be written in the unified form
\begin{equation}
	\hat{K}(\tau) = \exp(-\tau \mathcal{G}),
	\label{eq:17}
\end{equation}
where the generator $\mathcal{G}$ specifies the dynamical regime.

\subsection{Analytic Continuation}
To link and restore the quantum information between the real-time and imaginary-time Schrödinger equations (Eqs.~\eqref{eq:18} and ~\eqref{eq:19}, we define the Fourier transforms:
\begin{equation}
	\tilde{\psi}_{\mathrm{real}}(\mathbf{r},\omega)
	= \int_{-\infty}^{\infty} \psi_{\mathrm{real}}(\mathbf{r},t)e^{-i\omega t}\,dt,
\end{equation}
\begin{equation}
	\tilde{\psi}_{\mathrm{imag}}(\mathbf{r},\Omega)
	= \int_{0}^{\infty} \psi_{\mathrm{imag}}(\mathbf{r},\tau)e^{-i\Omega \tau}\,d\tau.
\end{equation}

The frequency-domain form of the real-time Schrödinger equation is
\begin{equation}
	(\hat{H}-\hbar\omega)\tilde{\psi}_{\mathrm{real}}(\mathbf{r},\omega)
	= \tilde{S}(\mathbf{r},\omega),
\end{equation}
while the imaginary-time counterpart becomes
\begin{equation}
	(\hat{H}+i\hbar\Omega)\tilde{\psi}_{\mathrm{imag}}(\mathbf{r},\Omega)
	= \tilde{S}(\mathbf{r},\Omega).
\end{equation}

We introduce the analytic continuation
\begin{equation}
	\Omega = i\omega,
\end{equation}
which yields the mapping $i\hbar\Omega = -\hbar\omega$. Therefore,
\begin{equation}
	\tilde{\psi}_{\mathrm{imag}}(\mathbf{r},\Omega)
	= \tilde{\psi}_{\mathrm{real}}(\mathbf{r}, i\omega).
\end{equation}

\subsection{Time-Domain Mapping}
Transforming back to the time domain, we obtain
\begin{equation}
	\psi_{\mathrm{imag}}(\mathbf{r},\tau)
	= \frac{1}{2\pi}\int_{-\infty}^{\infty}
	\tilde{\psi}_{\mathrm{real}}(\mathbf{r}, i\Omega)\, e^{i\Omega \tau}\, d\Omega.
\end{equation}

Equivalently, this defines a kernel representation
\begin{equation}
	\psi_{\mathrm{imag}}(\mathbf{r},\tau)
	= \int \psi_{\mathrm{real}}(\mathbf{r},t')\, K(\tau,t')\, dt'.
\end{equation}
where the kernel is given by
\begin{equation}
	K(\tau,t')
	= \frac{1}{2\pi}
	\int_{-\infty}^{\infty}
	e^{-\omega \tau} e^{i\omega t'}\, d\omega.
\end{equation}
More precisely, the kernel should be interpreted as a tempered distribution acting on a suitable space of test functions (e.g., Schwartz space), ensuring convergence after appropriate regularization.
This expression defines an analytic continuation structure in the complex frequency plane, mapping unitary evolution to contractive dynamics. However, the integral is not well-defined for large frequencies and must be understood in a distributional sense; without regularization, the kernel reduces to a generalized continuation of the Dirac delta distribution and lacks direct physical meaning.

\subsection{Square-Root Spectral Deformation and Fractional Semigroup Structure}
To obtain a mathematically well-defined mapping between real-time and imaginary-time dynamics, we formulate the transformation at the operator level using spectral calculus. Let $A = i\partial_t$ acting on \(L^2(\mathbb{R})\) with domain $D(A)=H^1(\mathbb{R})$. The operator \(A\) is self-adjoint and admits the spectral representation 
\begin{equation}
 A=\int_{\mathbb{R}} \omega\, dE_\omega
\end{equation}
where \(\{E_\omega\}\) denotes the associated projection-valued measure. This allows one to define functions of \(A\) via the functional calculus.
We introduce a square-root spectral deformation by defining the operator
\begin{equation}
\sqrt{A}
=
\int_{\mathbb{R}} \sigma(\omega)\, dE_\omega,
\end{equation}
where \(\sigma(\omega)\) is a fixed branch of the complex square root satisfying $\Re\,\sigma(\omega)\ge 0$. The introduction of the square-root spectral deformation is not merely a formal application of functional calculus, but reflects an intrinsic extremal property of the transformation between unitary and contractive dynamics.
Among all fractional spectral transformations of the form $A^\alpha$ ($0<\alpha\le1$), the square-root case $\alpha=1/2$ plays a distinguished role: it represents the minimal deformation that converts a unitary generator into a contractive semigroup while preserving spectral ordering and maintaining sensitivity to high-frequency components.
In this sense, the square-root generator defines a critical boundary between reversible (unitary) and strongly dissipative dynamics. It realizes a fractional evolution that is neither phase-preserving nor fully diffusive, but instead exhibits a scale-dependent attenuation governed by $\exp(-\tau \sqrt{\omega})$.
This identifies an intermediate universality class of quantum dynamics characterized by fractional spectral filtering and nonlocal temporal structure. The square-root deformation thus provides a canonical bridge between real-time and imaginary-time evolution, extending Wick rotation from a discrete mapping to a continuous spectral flow.
A canonical choice is
\begin{equation}
\sigma(\omega)=
\begin{cases}
	\sqrt{\omega}, & \omega \ge 0, \\
	i\sqrt{|\omega|}, & \omega < 0,
\end{cases}
\end{equation}
which ensures that the resulting operator generates a contractive semigroup.
We then define the evolution operator
\begin{equation}
	\hat K(\tau)
	=
	\exp\!\big(-\tau \sqrt{\hbar A}\big),
	\tag{21}
\end{equation}
which is well-defined through the functional calculus. By construction,
\(\{\hat K(\tau)\}_{\tau\ge 0}\) forms a strongly continuous semigroup satisfying
\begin{equation}
\hat K(\tau_1+\tau_2)
=
\hat K(\tau_1)\hat K(\tau_2),
\end{equation}
and
\begin{equation}
\|\hat K(\tau)\psi\|_2
\le
\|\psi\|_2.
\end{equation}
In the frequency domain, the action of \(\hat K(\tau)\) is diagonal:
\begin{equation}
\widetilde{\psi}(\omega)
\mapsto
\exp\!\big(-\tau \sqrt{\hbar}\,\sigma(\omega)\big)\,
\widetilde{\psi}(\omega),
\end{equation}
which shows explicitly that high-frequency components are exponentially suppressed according to the square-root of the spectral parameter.
This construction can be viewed as a generalized analytic continuation. 
While Wick rotation maps a unitary group to a contractive semigroup 
via $t \to -i\tau$, the present framework introduces a continuous 
spectral transformation at the level of the generator. 
In this sense, the square-root deformation interpolates between 
real-time and imaginary-time dynamics, extending Wick rotation 
from a discrete mapping to a spectral flow.
This defines a distinct dynamical universality class: whereas unitary 
evolution preserves spectral amplitudes and diffusive dynamics 
suppresses them exponentially, the square-root generator induces a 
fractional filtering mechanism governed by $\exp(-\tau \sqrt{\omega})$, 
intermediate between coherent oscillation and dissipative decay.
The operator $\sqrt{A}$ can be interpreted as a fractional power of 
$A=i\partial_t$, placing the construction within the general framework 
of fractional evolution equations and nonlocal dynamics.
A useful representation of the semigroup \(\hat K(\tau)\) is provided by the subordination formula
\begin{equation}
	\hat K(\tau)
	=
	\frac{1}{2\sqrt{\pi}}
	\int_0^\infty
	\frac{\tau}{s^{3/2}}
	\exp\!\left(-\frac{\tau^2}{4s}\right)
	e^{-s \hbar A}
	\, ds,
\end{equation}
which expresses the square-root evolution as a continuous superposition 
of standard exponential semigroups. This representation reveals that 
the fractional dynamics arises from a subordination mechanism, in which 
the evolution is governed by a distribution of diffusive time scales.
The corresponding representation has a natural physical interpretation: 
it describes a stochastic time-change process of Lévy type, where the 
effective evolution time becomes a random variable. In this sense, the 
square-root generator realizes a generalized diffusion with nonlocal 
temporal correlations. This interpretation connects analytic continuation to fractional kinetics and stochastic time-change processes, providing a physical basis for the square-root generator.
Formally, one may express $\hat K(\tau)$ in the time domain via the inverse Fourier transform,
\begin{equation}
	K(\tau,t')
	=
	\frac{1}{2\pi}
	\int_{\mathbb{R}}
	\exp\!\big(-\tau \sqrt{\hbar}\,\sigma(\omega)\big)
	e^{i\omega t'}\, d\omega,
\end{equation}
A mathematically controlled construction is obtained via the spectral theorem and functional calculus, which avoids ambiguities associated with direct evaluation of the oscillatory integral representation. This leads to a nonlocal kernel representation
\begin{equation}
(\hat K\psi)(\tau)
=
\int_{\mathbb{R}} K(\tau,t') \psi(t')\,dt'.
\end{equation}

However, this integral is generally oscillatory and must be understood in a distributional sense; a rigorous construction follows from the spectral definition above rather than direct contour integration. The square-root construction is rigorously justified when the generator is sectorial (in particular for non-negative self-adjoint operators), ensuring that the fractional power is well defined via holomorphic functional calculus.

\section{Spectral Structure of Inverse Reconstruction}
The inverse transformation, which formally reconstructs real-time dynamics from imaginary-time data, can be written as
\begin{equation}
	\psi_{\mathrm{real}}(\mathbf{r},t)
	=
	\exp\left(+t\sqrt{-\hbar \frac{\partial}{\partial \tau}}\right)
	\psi_{\mathrm{imag}}(\mathbf{r},\tau).
	\label{eq:47}
\end{equation}

In contrast to the forward evolution, the inverse operator is not contractive. Its spectral action amplifies high-frequency components that were exponentially suppressed under imaginary-time evolution, leading to an ill-conditioned reconstruction problem. As a result, the inverse mapping is highly sensitive to noise and perturbations, and becomes unbounded in the full spectral domain.

However, this instability is not uniform across all spectral modes. In practice, the exponential damping primarily affects high-frequency components, while low-energy modes remain relatively stable over finite imaginary times. This leads to a scale-dependent structure of invertibility: although exact reconstruction is unstable, partial recovery of the dynamics remains feasible within a well-defined spectral bandwidth.

From an operator-theoretic perspective, this establishes a qualified asymmetry between forward and inverse evolution. Imaginary-time propagation defines a contractive semigroup acting as a spectral low-pass filter, whereas the inverse transformation corresponds to a non-contractive operator that realizes a bandwidth-limited reconstruction rather than a fully unstable inversion.

The mapping between real-time and imaginary-time dynamics can therefore be understood as a nonlocal operator acting on the temporal domain, with kernel
\begin{equation}
	\exp\left(-\tau\sqrt{-\hbar \frac{\partial}{\partial t}}\right),
\end{equation}
revealing that imaginary-time evolution implements a square-root spectral filter. Within this framework, the limits of reconstruction are governed not by effective irreversibility under finite noise and resolution, but by the extent to which spectral information survives this filtering and can be stably recovered.

The instability of inverse reconstruction can be characterized by a sharp spectral bound. 
Consider imaginary-time data of the form
\begin{equation}
	y_n = e^{-E_n \tau} x_n + \delta_n,
\end{equation}
where $\delta_n$ denotes noise with amplitude $\|\delta\| \le \epsilon$. 
A reconstruction is deemed stable if perturbations remain bounded under inversion, 
$| \delta x_n | \lesssim \mathcal{O}(1)$. 
Since inversion amplifies spectral components as $e^{E_n \tau}$, this condition requires
\begin{equation}
	e^{E_n \tau} \epsilon \lesssim 1.
\end{equation}
This defines a critical energy scale
\begin{equation}
	E_c(\tau, \epsilon) \sim \frac{1}{\tau} \log \frac{1}{\epsilon}.
\end{equation}
We thus obtain a recoverability bound: modes with $E_n < E_c$ are stably reconstructable, 
whereas modes with $E_n > E_c$ are exponentially unstable under inversion. 
This establishes a sharp, scale-dependent spectral boundary separating a recoverable low-energy sector from an irrecoverable high-frequency regime.
Importantly, this bound is independent of the detailed structure of the underlying system and depends only on the spectral parameter $E_n$, the imaginary time $\tau$, and noise level $\epsilon$, demonstrating that analytic continuation is intrinsically constrained by a bandwidth-limited structure.

The inverse reconstruction problem can be formulated within the framework of compact operators. In an appropriate spectral representation, the forward map takes the form
\begin{equation}
y(\omega) = K_\tau(\omega) x(\omega), \qquad K_\tau(\omega) = e^{-\tau \omega},
\end{equation}
which corresponds to a Laplace-type transformation.
When viewed as an operator between suitable Hilbert spaces (e.g., weighted $L^2$ spaces), $K_\tau$ defines a compact operator whose singular values decay rapidly with frequency. This compactness is the fundamental source of ill-posedness.
More explicitly, expanding in a suitable orthonormal basis $\{ \phi_n \}$, the operator admits a singular value decomposition
\begin{equation}
K_\tau = \sum_{n} \sigma_n \langle \cdot, v_n \rangle u_n,
\end{equation}
where the singular values satisfy an asymptotic decay of the form $\sigma_n \sim e^{-c\,\tau\, n}$, for some constant $c>0$ depending on the spectral parametrization. This exponential decay implies that inversion amplifies high-index components as $\sigma_n^{-1} \sim e^{c \tau n}$.

A necessary condition for stable inversion is the Picard condition, which requires that the expansion coefficients of the data, $\langle y, u_n \rangle$, decay at least as fast as the singular values $\sigma_n$. In the present setting, however, noise introduces a flat or slowly decaying component, violating this condition at high indices. As a result, the inversion becomes dominated by noise amplification.
In the presence of noise $\| \delta y \| \le \epsilon$, the reconstructed coefficients behave as
\begin{equation}
\delta x_n \sim \frac{1}{\sigma_n} \delta y_n \sim e^{c \tau n} \delta y_n,
\end{equation}
leading to exponential growth of errors in high-frequency components. Regularization (e.g., Tikhonov) effectively truncates this growth by suppressing modes with small singular values. The resulting reconstruction error exhibits a trade-off between bias and variance, consistent with classical results for exponentially ill-posed problems.
The exponential decay of singular values provides an operator-theoretic basis for the recoverability bound derived above. The effective cutoff scale $E_c(\tau,\epsilon)$ corresponds to the index at which noise amplification becomes dominant, yielding a bandwidth-limited reconstruction regime.

\begin{figure*}[t]
	\centering
	\includegraphics[width=\textwidth]{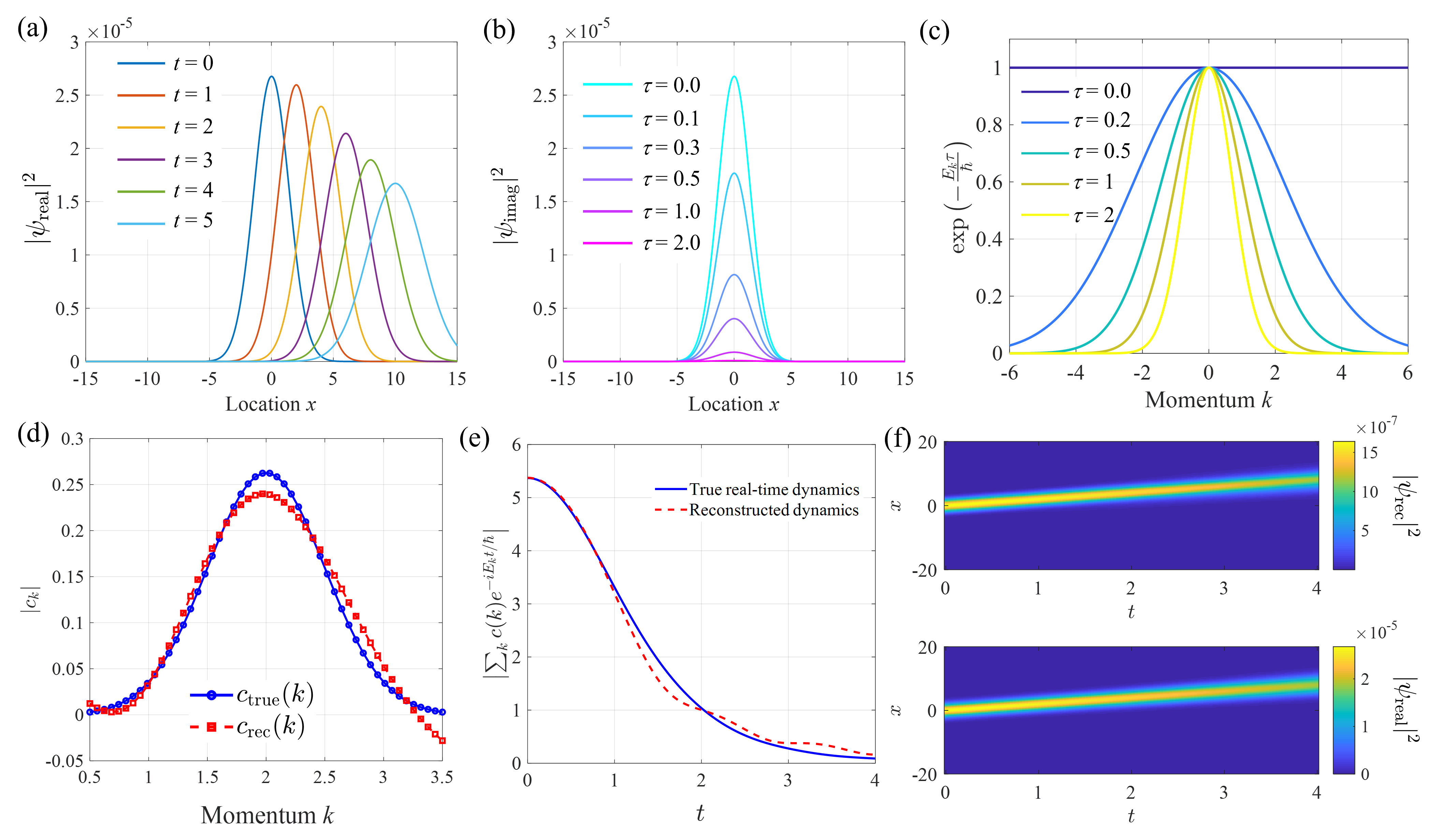}
	\caption{Real- and imaginary-time dynamics of a free particle and inverse reconstruction. (a) Real-time evolution of a free-particle wave packet. (b) Imaginary-time evolution of a free-particle wave packet. (c) Momentum-space decay factor under imaginary-time evolution. (d) Reconstruction of the spectral coefficients while suppressing high-frequency instability. (e) Reconstructed real-time dynamics from imaginary-time data. (f) Comparison between reconstructed and exact spatiotemporal dynamics.}\label{fig1}
\end{figure*}

\section{Quantum Dynamical Systems}
\subsection{Schrödinger Equation}
As a concrete realization of the general generator framework, we now consider the Schrödinger equation, where the generator is given by $\mathcal{G} = \frac{i}{\hbar}\hat{H}$. The real-time evolution satisfies
\begin{equation}
	i\hbar \frac{\partial \psi_{\mathrm{real}}(\mathbf{r},t)}{\partial t}
	= \hat{H}\psi_{\mathrm{real}}(\mathbf{r},t) + S(\mathbf{r},t),
	\label{eq:18}
\end{equation}
where $\hat{H}$ is the Hamiltonian operator and $\hbar$ is the reduced Planck constant. The evolution is unitary, and the solution exhibits oscillatory phase factors of the form $e^{-iEt/\hbar}$.

Based on Wick rotation $t \rightarrow -i\tau$, we obtain the imaginary-time dynamics:
\begin{equation}
	\frac{\partial \psi_{\mathrm{imag}}(\mathbf{r},\tau)}{\partial \tau}
	= -\frac{1}{\hbar}\hat{H}\psi_{\mathrm{imag}}(\mathbf{r},\tau)
	+ S(\mathbf{r},\tau).
	\label{eq:19}
\end{equation}
Expanding the wave function in the energy eigenbasis, we write
\begin{equation}
	\psi = \sum_n c_n \phi_n.
\end{equation}
The real-time and imaginary-time solutions take the forms
\begin{equation}
	\psi_{\mathrm{real}}(t)
	= \sum_n c_n e^{-iE_n t/\hbar} \phi_{\mathrm{real},n},
\end{equation}
\begin{equation}
	\psi_{\mathrm{imag}}(\tau)
	= \sum_n c_n e^{-E_n \tau/\hbar} \phi_{\mathrm{imag},n},
\end{equation}
showing the transition from unitary oscillation to contractive decay. In the long-time limit, the system naturally converges to the ground state,
\begin{equation}
	\psi_{\mathrm{imag}}(\tau \to \infty) \to \phi_0.
\end{equation}

\begin{figure*}[t]
	\centering
	\includegraphics[width=\textwidth]{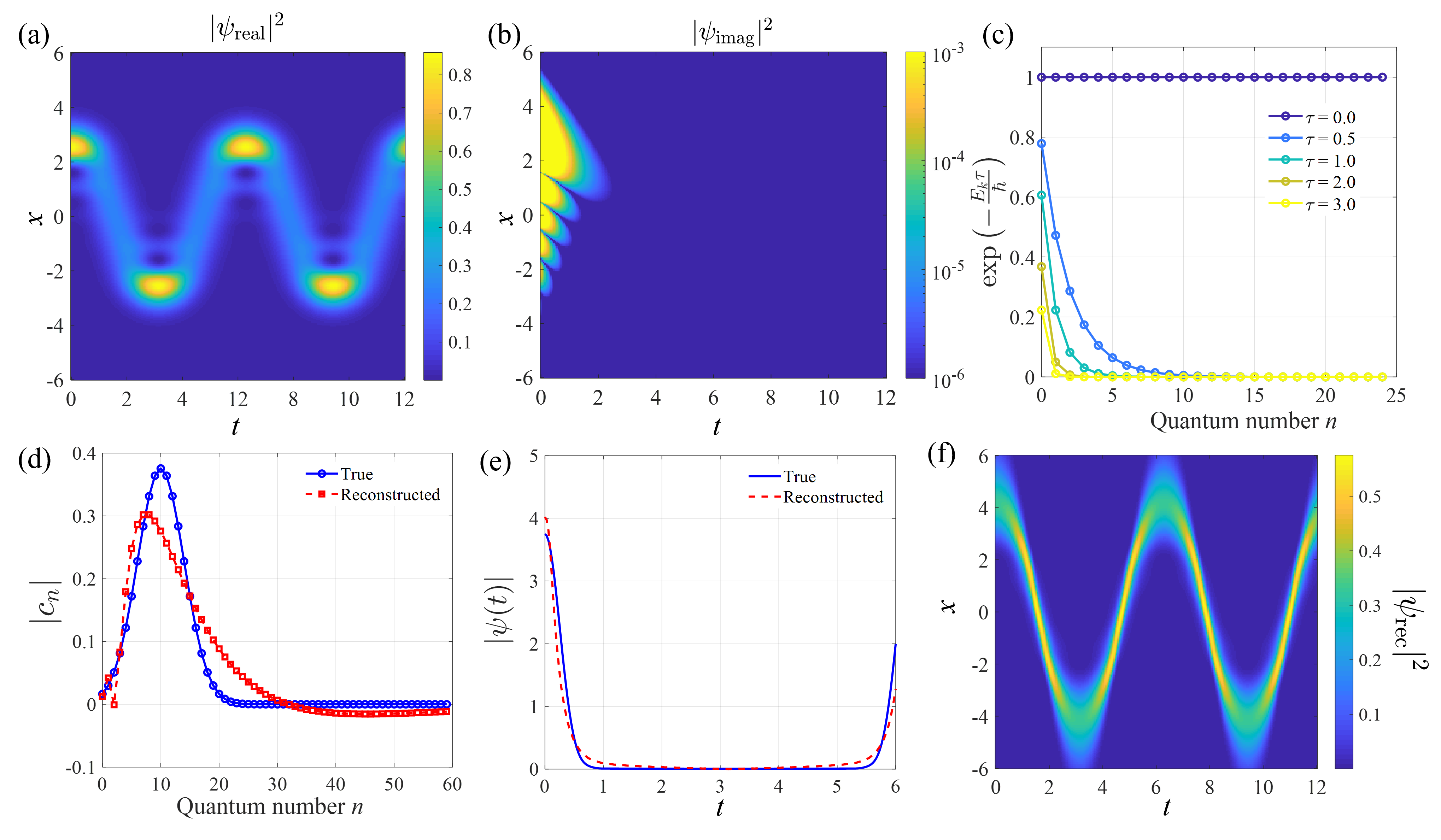}
	\caption{Discrete-spectrum imaginary-time filtering and inverse reconstruction in the quantum harmonic oscillator. (a) Real-time spatiotemporal dynamics of the harmonic oscillator. (b) Imaginary-time spatiotemporal dynamics of the harmonic oscillator. (c) Exponential spectral damping under imaginary-time evolution in the harmonic oscillator. (d) Spectral reconstruction in the harmonic oscillator. (e) Reconstruction of real-time dynamics from imaginary-time data in the harmonic oscillator. (f) Reconstructed spatiotemporal dynamics of the harmonic oscillator.}\label{fig2}
\end{figure*}

\subsection{Continuous and Discrete Spectra}
We first consider a one-dimensional free particle with Hamiltonian
\begin{equation}
	\hat{H} = -\frac{\hbar^2}{2m}\partial_x^2.
\end{equation}

The energy eigenstates are plane waves
\begin{equation}
	\phi_k(x) = \frac{1}{\sqrt{2\pi}} e^{ikx},
	\qquad
	E_k = \frac{\hbar^2 k^2}{2m}.
\end{equation}

Real-time evolution reads
\begin{equation}
	\psi_{\mathrm{real}}(x,t)
	=
	\int_{-\infty}^{\infty}
	dk\;
	\tilde{\psi}_0(k)\,e^{ikx - iE_k t/\hbar},
\end{equation}
which preserves the momentum distribution and encodes dynamics purely through phase rotation. Under Wick rotation \(t \to -i\tau\),
\begin{equation}
	\psi_{\mathrm{imag}}(x,\tau)
	=
	\int dk\;
	\tilde{\psi}_0(k)\,e^{ikx}\,e^{-\tau E_k/\hbar},
\end{equation}
with
\begin{equation}
	e^{-\tau E_k/\hbar}
	=
	\exp\!\left(-\frac{\hbar k^2}{2m}\tau\right).
\end{equation}

Thus, imaginary-time evolution induces a Gaussian suppression in momentum space.

We next consider the harmonic oscillator,
\begin{equation}
	\hat{H}
	=
	\hbar\omega\left(a^\dagger a + \frac{1}{2}\right),
\end{equation}
with eigenstates \(|n\rangle\) and energies
\begin{equation}
	E_n = \hbar\omega\left(n + \frac{1}{2}\right),
	\qquad n = 0,1,2,\dots
\end{equation}

An arbitrary initial state admits the expansion
\begin{equation}
	\psi_{\mathrm{real}}(t)
	=
	\sum_{n=0}^{\infty}
	c_n e^{-iE_n t/\hbar}\,|n\rangle.
\end{equation}

Imaginary-time evolution yields
\begin{equation}
	\psi_{\mathrm{imag}}(\tau)
	=
	\sum_{n=0}^{\infty}
	c_n e^{-E_n \tau/\hbar}\,|n\rangle.
\end{equation}

As \(\tau \to \infty\),
\begin{equation}
	\psi_{\mathrm{imag}}(\tau)
	\;\longrightarrow\;
	c_0 e^{-E_0 \tau/\hbar}|0\rangle,
\end{equation}
demonstrating explicit ground-state projection.

\begin{figure*}[t]
	\centering
	\includegraphics[width=\textwidth]{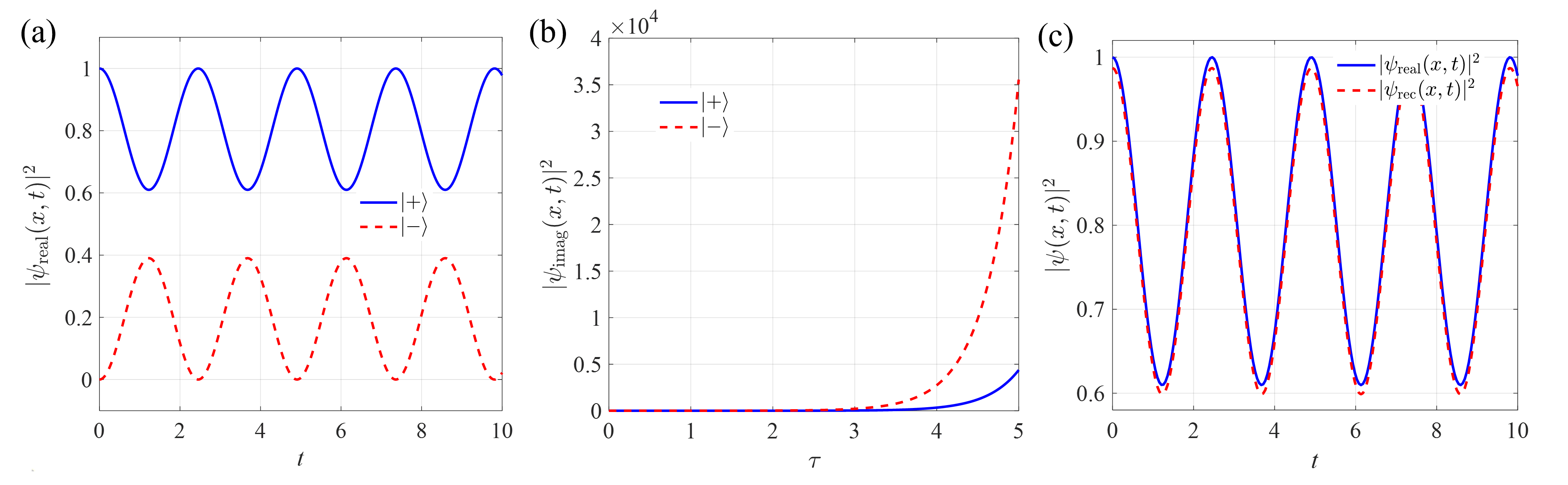}
	\caption{Dynamics of a two-level system under real- and imaginary-time evolution. (a) Real-time Rabi oscillations. (b) Imaginary-time projection onto the ground state. (c) Reconstructed real-time population dynamics from imaginary-time data.}\label{fig3}
\end{figure*}

\subsection{Quantum-Coherent Dynamics}
We first consider the minimal quantum system exhibiting genuine phase-dependent dynamics: a two-level system described by the Hamiltonian
\begin{equation}
	\hat{H}_0 = \frac{\Delta}{2}\,\sigma_z,
\end{equation}
where \(\sigma_z\) is the Pauli matrix and \(\Delta\) denotes the level splitting.

An arbitrary initial state
\begin{equation}
	|\psi(0)\rangle = c_+|+\rangle + c_-|-\rangle
\end{equation}
evolves in real time as
\begin{equation}
	|\psi_{\mathrm{real}}(t)\rangle
	=
	c_+ e^{-i\Delta t/2\hbar}|+\rangle
	+
	c_- e^{+i\Delta t/2\hbar}|-\rangle.
\end{equation}

The dynamics corresponds to a coherent rotation on the Bloch sphere about the \(z\)-axis, with all physical observables depending explicitly on the relative phase between the two amplitudes. Under Wick rotation, the imaginary-time Schrödinger equation reads
\begin{equation}
	\partial_\tau |\psi_{\mathrm{imag}}(\tau)\rangle
	=
	-\frac{1}{\hbar}\hat{H}_0|\psi_{\mathrm{imag}}(\tau)\rangle,
\end{equation}
with solution
\begin{equation}
	|\psi_{\mathrm{imag}}(\tau)\rangle
	=
	c_+ e^{-\Delta\tau/2\hbar}|+\rangle
	+
	c_- e^{+\Delta\tau/2\hbar}|-\rangle.
\end{equation}

For \(\Delta > 0\), the excited component is exponentially suppressed, and the state is rapidly projected onto the ground-state manifold. Importantly, the complex phase structure present in Eq.~(71) is entirely eliminated. To further enhance phase sensitivity, we consider a driven Hamiltonian
\begin{equation}
	\hat{H}(t)
	=
	\frac{\Delta}{2}\sigma_z + \Omega \sigma_x,
\end{equation}
which generates nontrivial trajectories on the Bloch sphere involving population transfer and dynamical interference. In real time, the evolution exhibits Rabi oscillations and phase-dependent beating. In imaginary time, however, the propagator
\begin{equation}
	\hat{K}(\tau)
	=
	\exp\!\left[-\tau(\Delta\sigma_z/2 + \Omega\sigma_x)/\hbar\right]
\end{equation}
defines a matrix-valued contraction semigroup that suppresses both oscillatory behavior and transverse coherence. This model allows one to explicitly quantify the extent to which phase information is irreversibly lost under imaginary-time propagation. We reconstruct the real-time dynamics from imaginary-time data using \(\ell_2\)-regularized inversion in the energy (or Pauli) basis:
\begin{equation}
	\min_{\mathbf{c}}
	\;
	\|K(\tau)\mathbf{c} - \mathbf{y}\|_2^2
	+
	\lambda \|\mathbf{c}\|_2^2,
\end{equation}
where \(\mathbf{c} = (c_+, c_-)\) denotes the spectral coefficients and \(\mathbf{y}\) represents the imaginary-time state. The recovered real-time wave function is then evolved unitarily and compared with the exact solution.

We next consider a particle in a finite potential landscape,
\begin{equation}
	\hat{H}
	=
	-\frac{\hbar^2}{2m}\partial_x^2 + V(x),
\end{equation}
where \(V(x)\) is chosen as a symmetric double-well potential. The low-energy spectrum consists of nearly degenerate symmetric and antisymmetric states,
\begin{equation}
	E_{\pm} = E_0 \pm \frac{\delta E}{2},
\end{equation}
with exponentially small splitting \(\delta E\). A localized wave packet initially trapped in one well can be expressed as a superposition of the two lowest eigenstates. Real-time evolution yields coherent tunneling oscillations:
\begin{equation}
	\psi_{\mathrm{real}}(t)
	\sim
	e^{-iE_+ t/\hbar}\phi_+
	+
	e^{-iE_- t/\hbar}\phi_-,
\end{equation}
with tunneling frequency
\begin{equation}
	\omega_{\mathrm{tun}} = \frac{\delta E}{\hbar}.
\end{equation}

Imaginary-time evolution gives
\begin{equation}
	\psi_{\mathrm{imag}}(\tau)
	=
	e^{-E_+ \tau/\hbar}\phi_+
	+
	e^{-E_- \tau/\hbar}\phi_-,
\end{equation}
which rapidly suppresses the excited contribution and selects the ground-state configuration. As a result, tunneling oscillations are completely eliminated in the long-\(\tau\) limit. Applying \(\ell_2\)-regularized inversion in the low-energy eigenbasis, we test whether: (i) the tunneling frequency \(\omega_{\mathrm{tun}}\) can be recovered, and (ii) the wave-packet splitting and recombination dynamics can be reconstructed in real time. We find that partial recovery is possible only within a restricted range of imaginary time \(\tau\) and noise level. Beyond this regime, the spectral gap becomes too small relative to the exponential suppression, and reconstruction fails.

\begin{figure*}[t]
	\centering
	\includegraphics[width=\textwidth]{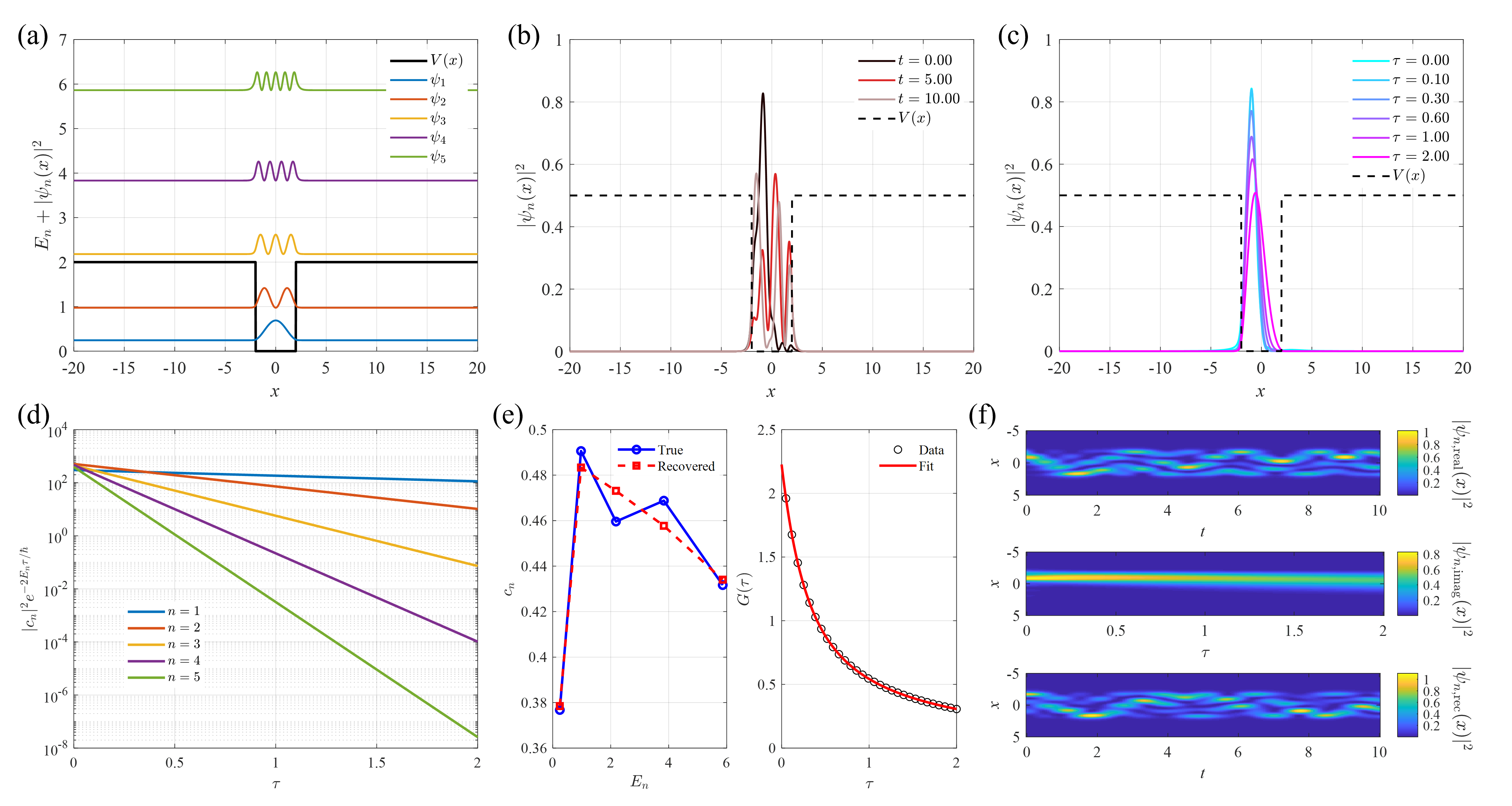}
	\caption{Finite potential well dynamics, imaginary-time filtering, and inverse spectral reconstruction. (a) Bound states in the finite potential well. (b) Real-time tunneling dynamics of an initially localized wave packet. (c) Imaginary-time evolution in the finite potential well. (d) Spectral decay under imaginary-time evolution. (e) Spectral reconstruction from imaginary-time data (left) and fit of imaginary-time data using reconstructed spectral coefficients (right). (f) Comparison between real-time, imaginary-time, and reconstructed finite potential well dynamics.}\label{fig4}
\end{figure*}

\subsection{Non-Hermitian and Open Quantum Systems}
We consider quantum systems governed by an effective non-Hermitian Hamiltonian of the form
\begin{equation}
	\hat{H} = \hat{H}_0 - i\gamma \hat{W},
\end{equation}
where \(\hat{H}_0 = \hat{H}_0^\dagger\) describes coherent dynamics, \(\hat{W} = \hat{W}^\dagger \ge 0\) represents loss or decay channels, and \(\gamma > 0\) controls the strength of dissipation. Such Hamiltonians arise naturally in a wide range of physical contexts, including open quantum systems, photonic platforms with absorption, and effective descriptions of measurement backaction. The real-time Schrödinger equation,
\begin{equation}
	i\hbar \partial_t \psi = \hat{H}\psi,
\end{equation}
is no longer norm preserving. Instead, the evolution generator has a spectrum
\begin{equation}
	\lambda_n = E_n - i\Gamma_n,
\end{equation}
with \(\Gamma_n \ge 0\). Formally, both non-Hermitian real-time evolution and Hermitian imaginary-time evolution are governed by contraction semigroups:
\begin{equation}
	\psi(t) = e^{-i\hat{H} t/\hbar}\psi(0),
	\qquad
	\psi(\tau) = e^{-\hat{H}_0 \tau/\hbar}\psi(0).
\end{equation}

In both cases, the real parts of the generator spectrum control exponential decay, while imaginary components generate residual oscillations.

Non-Hermitian generators generically support exceptional points (EPs), at which both eigenvalues and eigenvectors coalesce. Near an EP of order two, the generator can be reduced to the canonical form
\begin{equation}
	\mathcal{G}_{\mathrm{EP}}
	=
	\begin{pmatrix}
		\lambda & 1 \\
		0 & \lambda
	\end{pmatrix}.
\end{equation}

The associated evolution operator contains polynomial prefactors,
\begin{equation}
	e^{-\tau \mathcal{G}_{\mathrm{EP}}}
	=
	e^{-\lambda \tau}
	\begin{pmatrix}
		1 & -\tau \\
		0 & 1
	\end{pmatrix},
\end{equation}
indicating anomalous temporal scaling. Applying \(\ell_2\)-regularized reconstruction to imaginary-time or non-Hermitian evolution data reveals a sharp degradation of performance near exceptional points. Specifically, we observe that: (i) the effective spectral representation becomes increasingly non-orthogonal due to eigenvector coalescence; (ii) the condition number of the generator diverges algebraically near the EP; (iii) small perturbations are amplified polynomially rather than exponentially. As a result, the optimization problem
\begin{equation}
	\min_x \|Kx - y\|_2^2 + \lambda \|x\|_2^2
\end{equation}
exhibits strong sensitivity to noise, even when the forward map remains contractive. To quantify the limits of reversibility, we analyze the reconstruction error as a function of the spectral condition number \(\kappa(\mathcal{G})\) of the generator. Numerically, we find
\begin{equation}
	\varepsilon_{\mathrm{rec}}
	\sim
	\kappa(\mathcal{G})\,e^{-\tau \Delta \Re \lambda},
\end{equation}
where \(\Delta \Re \lambda\) denotes the minimal spectral gap in the real part of the eigenvalues. This scaling demonstrates that irreversibility is governed not only by spectral damping but also by non-normal amplification intrinsic to non-Hermitian operators.

\begin{figure*}[t]
	\centering
	\includegraphics[width=\textwidth]{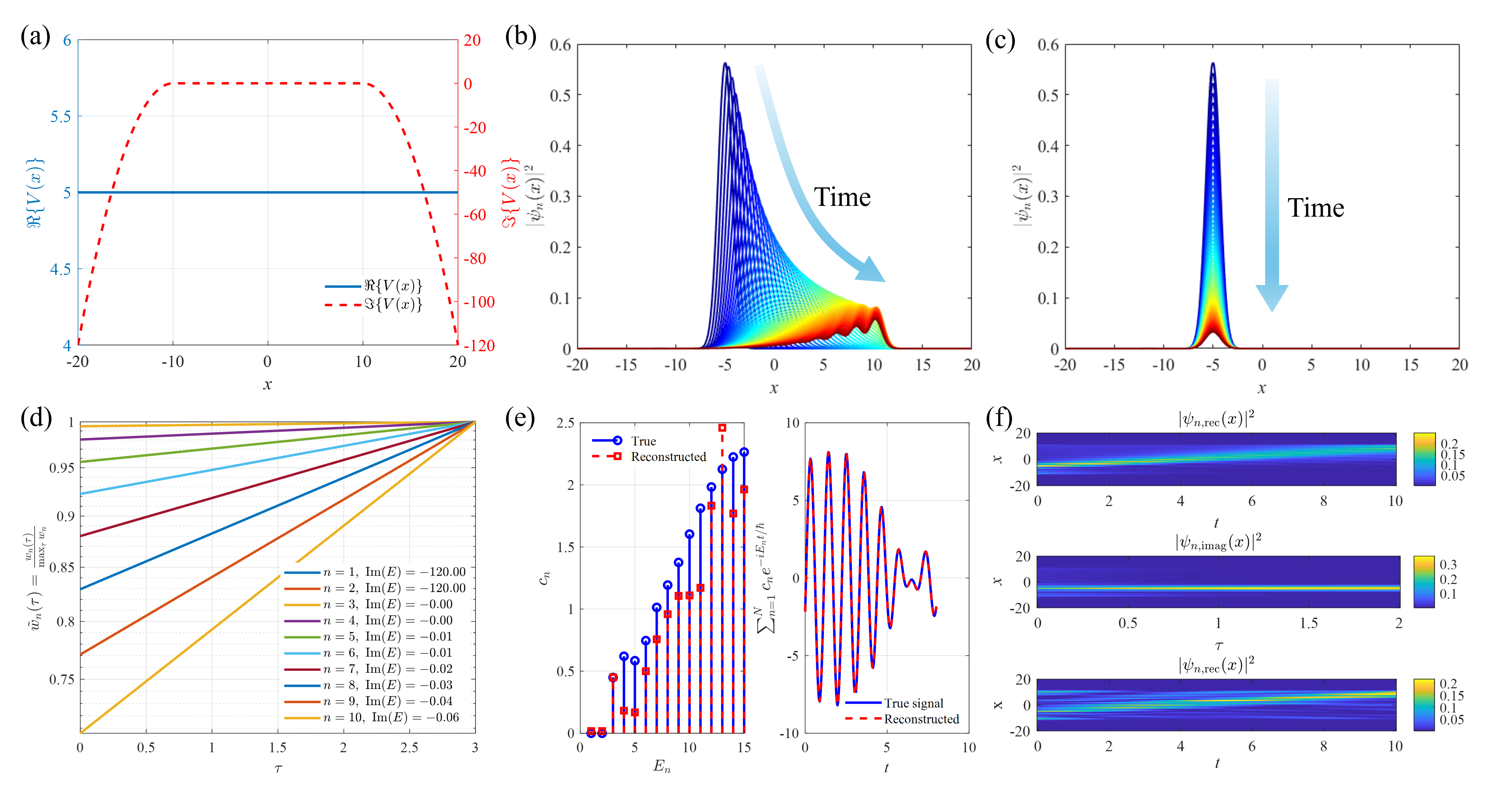}
	\caption{Non-Hermitian dynamics, spectral instability, and inverse reconstruction. (a) Non-Hermitian complex potential. (b) Non-Hermitian real-time dynamics of a wave packet. (c) Imaginary-time-like evolution in a non-Hermitian system. (d) Spectral dynamics under non-Hermitian evolution. (e) Reconstruction of spectral coefficients (left) and reconstructed forward signal (right). (f) Comparison of real-time, imaginary-time, and reconstructed dynamics in a non-Hermitian system.}\label{fig5}
\end{figure*}

We now extend the framework to fully open quantum systems described by a Lindblad master equation,
\begin{equation}
	\dot{\rho}
	=
	-i[\hat{H},\rho]
	+
	\sum_k
	\left(
	L_k \rho L_k^\dagger
	-
	\frac{1}{2}\{L_k^\dagger L_k,\rho\}
	\right).
\end{equation}

In superoperator form, this reads
\begin{equation}
	\partial_\tau \rho(\tau)
	=
	-\mathcal{L}\rho(\tau),
	\qquad
	\rho(\tau) = e^{-\tau \mathcal{L}} \rho(0),
\end{equation}
where \(\mathcal{L}\) is the Lindbladian generator acting on operator space. The Lindblad superoperator defines a contractive semigroup on the space of density matrices. Its spectrum satisfies
\begin{equation}
	\Re\, \sigma(\mathcal{L}) \ge 0,
\end{equation}
ensuring convergence toward stationary states. From this viewpoint, Lindblad dynamics represents a generalized imaginary-time flow in Liouville space, with decay rates determined by dissipative channels. We investigate whether \(\ell_2\)-regularization can partially recover coherent unitary contributions embedded within dissipative trajectories. Specifically, we test whether: (i) oscillatory components associated with \([H,\rho]\) can be reconstructed, and (ii) decay rates associated with \(L_k\) channels impose fundamental recovery limits. The optimization problem generalizes to
\begin{equation}
	\min_X
	\;
	\|\mathcal{K}X - Y\|_2^2
	+
	\lambda \|X\|_2^2,
\end{equation}
where \(X\) denotes the Liouville-space representation of \(\rho(0)\).
In contrast to the self-adjoint case, non-Hermitian generators are generically non-normal, and therefore do not admit an orthogonal spectral decomposition. In such systems, the spectral properties alone are insufficient to characterize the dynamics: the geometry of the eigenvectors and the associated pseudospectrum play a crucial role.
In particular, the eigenvectors may become strongly non-orthogonal, requiring a biorthogonal representation, and the condition number of the spectral decomposition can grow rapidly near exceptional points. As a result, transient amplification and sensitivity to perturbations can occur even when the spectrum suggests contractive behavior.
The present framework captures these effects at the level of effective spectral conditioning, but a full treatment in terms of pseudospectral theory or Jordan block structure lies beyond the scope of this work.

\subsection{Relativistic Dirac Dynamics}
We finally consider relativistic quantum dynamics governed by the Dirac Hamiltonian
\begin{equation}
	\hat{H}_D
	=
	-i\hbar c\,\boldsymbol{\alpha}\cdot\nabla
	+
	\beta mc^2,
\end{equation}
acting on four-component spinor fields \(\Psi(\mathbf{r},t)\in\mathbb{C}^4\). Here \(\boldsymbol{\alpha}\) and \(\beta\) are Dirac matrices satisfying the Clifford algebra. The spectrum of \(\hat{H}_D\) is given by
\begin{equation}
	E_{\mathbf{p}}^{\pm}
	=
	\pm \sqrt{c^2 p^2 + m^2 c^4},
\end{equation}
corresponding to positive- and negative-energy branches separated by a mass gap \(2mc^2\). Dirac dynamics differs qualitatively from all previously considered cases: (i) The spectrum is unbounded from both above and below. (ii) Physical observables depend on interference between positive- and negative-energy components. (iii) The generator acts on both spatial and internal (spinor) degrees of freedom. These features make relativistic systems an intrinsically delicate testing ground for any attempt to recover real-time dynamics from contractive evolution.

Under Wick rotation, the imaginary-time evolution is governed by
\begin{equation}
	\partial_\tau \Psi(\tau)
	=
	-\frac{1}{\hbar}\hat{H}_D \Psi(\tau),
\end{equation}
with solution
\begin{equation}
	\Psi(\tau)
	=
	e^{-\tau \hat{H}_D/\hbar}\Psi(0).
\end{equation}

In the eigenbasis of \(\hat{H}_D\),
\begin{equation}
	\Psi(\tau)
	=
	\sum_{n,\pm}
	c_n^{\pm}
	e^{-E_n^{\pm}\tau/\hbar}
	u_n^{\pm}.
\end{equation}

For \(\tau>0\), positive-energy components decay exponentially, whereas negative-energy components grow, reflecting the indefinite nature of the Dirac spectrum. In practice, relativistic imaginary-time dynamics must be supplemented by a projection onto the positive-energy subspace, consistent with the standard no-pair approximation. The effective imaginary-time dynamics thus reads
\begin{equation}
	\Psi^{(+)}(\tau)
	=
	P_+ e^{-\tau \hat{H}_D/\hbar} P_+ \Psi(0),
\end{equation}
where \(P_+\) denotes the projector onto the positive-energy branch.
Relativistic quantum systems introduce additional structural features beyond the non-relativistic setting. In particular, the Dirac operator possesses an intrinsically indefinite spectrum with both positive- and negative-energy branches, and physical observables depend on their interference.
Under imaginary-time evolution, this spectral indefiniteness leads to fundamentally different behavior of positive- and negative-energy modes, including exponential growth in the latter. As a result, a consistent formulation requires projection onto a positive-energy subspace, analogous to the no-pair approximation.
From the perspective of the present framework, this projection introduces an additional source of information suppression, beyond spectral filtering. A full treatment would require addressing issues such as reflection positivity and Euclidean reconstruction, which lie outside the scope of this work.

\subsection{Inverse Reconstruction and Regularization}
We now consider the inverse problem of reconstructing real-time dynamics from imaginary-time data within the unified spectral framework developed above. For all classes of systems considered—ranging from Schrödinger dynamics and systems with continuous or discrete spectra, to quantum-coherent models, and non-Hermitian generators—the imaginary-time evolution can be written in the abstract form
\begin{equation}
	y = K(\tau)x,
	\qquad
	K(\tau) = \exp\!\left(-\frac{\tau}{\hbar}\hat{H}\right),
\end{equation}
where \(x\) denotes the real-time spectral coefficients (e.g., momentum modes, energy eigenstates, or spinor components), and \(y\) represents the corresponding imaginary-time data.

In the spectral basis of the generator, the operator \(K(\tau)\) acts diagonally as
\begin{equation}
	K(\tau): c_n \mapsto e^{-E_n \tau/\hbar} c_n,
\end{equation}
which exponentially suppresses high-energy components. While this forward evolution is contractive and well-posed, the inverse problem,
\begin{equation}
	x = K(\tau)^{-1} y,
\end{equation}
is intrinsically ill-posed: modes associated with large eigenvalues \(E_n\), or high frequencies in the continuous spectrum, are exponentially amplified. This instability is further exacerbated in non-Hermitian systems due to non-normal amplification.

To obtain a stable reconstruction, we introduce \(\ell_2\)-regularization in the spectral domain:
\begin{equation}
	\min_x
	\;
	\|K(\tau)x - y\|_2^2
	+
	\lambda \|x\|_2^2,
\end{equation}
where \(\lambda > 0\) is the regularization parameter. This formulation corresponds to Tikhonov regularization and can be interpreted as imposing a Gaussian prior on the spectral coefficients, penalizing high-energy contributions that are most susceptible to instability.

The solution admits the closed-form expression
\begin{equation}
	x = (K^\top K + \lambda I)^{-1} K^\top y,
\end{equation}
which defines a stable reconstruction operator. In spectral representation, this regularized inverse acts as a controlled reweighting,
\begin{equation}
	c_n \mapsto 
	\frac{e^{-E_n \tau/\hbar}}{e^{-2E_n \tau/\hbar} + \lambda}
	\, y_n,
\end{equation}
thereby suppressing the exponential amplification of high-energy modes.

This framework applies uniformly across all systems considered. For continuous spectra (free particle), the reconstruction operates in momentum space; for discrete systems (harmonic oscillator, few-level models), in the energy eigenbasis; and for non-Hermitian systems, in a biorthogonal spectral representation. In all cases, the interplay between spectral damping and regularization determines the recoverable information content.

We therefore conclude that while imaginary-time evolution defines a well-posed contractive semigroup, the inverse reconstruction is fundamentally unstable and can only be carried out in a regularized sense. The Tikhonov framework provides a unified and systematically controllable approach to approximate the real-time dynamics across all dynamical regimes considered in this work.

\section{Results and Discussion}
The numerical results presented here are intended to illustrate the spectral mechanisms identified above rather than to provide optimized reconstruction schemes. In particular, we do not aim to compete with established numerical approaches such as maximum entropy or Padé reconstruction, but instead use simple regularization to expose the intrinsic scaling behavior and recoverability structure. Quantitative benchmarking against optimized methods lies beyond the scope of this work and does not affect the generality of the spectral bounds derived here.

\subsection{Continuous and Discrete Spectra}
Figure~\ref{fig1} illustrates the contrast between unitary real-time propagation and contractive imaginary-time evolution, together with the performance of inverse reconstruction. In real time (Fig.~\ref{fig1}(a)), the free-particle wave packet exhibits coherent translation with dispersive broadening, reflecting phase-preserving dynamics. In imaginary time (Fig.~\ref{fig1}(b)), the amplitude is rapidly suppressed while the profile remains Gaussian, indicating the absence of phase propagation and the dominance of spectral filtering. This behavior is clearly seen in momentum space (Fig.~\ref{fig1}(c)), where the propagator introduces an exponential damping factor $\exp(-E_k \tau/\hbar)$, selectively suppressing high-momentum components and concentrating weight in low-energy modes.
The inverse reconstruction (Figs.~\ref{fig1}(d)-(f)) demonstrates that the dominant spectral features can be faithfully recovered. Tikhonov regularization stabilizes the inversion and accurately reconstructs the low-energy spectral envelope, which governs the global structure of the dynamics. Deviations become noticeable only at large momenta, where strong damping amplifies sensitivity to noise. Nevertheless, the reconstructed dynamics reproduces the main propagation characteristics, including the overall spreading trend, as observed in both the temporal evolution (Fig.~\ref{fig1}(e)) and the spatiotemporal comparison (Fig.~\ref{fig1}(f)).
These results indicate that imaginary-time evolution acts as a compressive spectral filter, effectively concentrating information into a reduced low-energy subspace. Within this subspace, the reconstruction is robust and quantitatively reliable, while outside it the recoverable information becomes progressively limited. The resulting asymmetry between forward propagation and inverse recovery therefore reflects a scale-dependent resolution limit rather than a fundamental breakdown of invertibility.
To quantify the reconstruction accuracy, we observe that the deviation between reconstructed and exact dynamics follows the scaling form
\begin{equation}
	\epsilon_{\mathrm{rec}} \sim e^{-\tau \Delta E},
\end{equation}
where $\Delta E$ denotes the effective spectral gap of the dominant modes. This scaling holds across both continuous and discrete spectra, indicating a universal relation between spectral filtering and reconstruction accuracy. Combined with the logarithmic recoverability bound, this identifies an effective spectral window that narrows with imaginary time, providing a quantitative relation between evolution duration, spectral gap, and achievable reconstruction precision.

Figure~\ref{fig2} illustrates spectral filtering by imaginary-time evolution in a discrete system, using the quantum harmonic oscillator as a prototype. In real time (Fig.~\ref{fig2}(a)), the wave packet exhibits periodic, phase-coherent oscillations reflecting the equally spaced spectrum. In contrast, imaginary-time evolution (Fig.~\ref{fig2}(b)) suppresses oscillations and drives the system toward the ground state, consistent with contractive dynamics.
The filtering mechanism is explicit at the spectral level (Fig.~\ref{fig2}(c)): each eigenmode decays as $\exp(-E_n \tau/\hbar)$, with higher-energy states suppressed more rapidly. This leads to progressive truncation of the effective spectral bandwidth, resulting in a mode-by-mode concentration onto low-energy components.
The impact on inverse reconstruction is shown in Figs.~\ref{fig2}(d)-(f). While low-energy coefficients can be stably recovered, higher-energy modes become strongly biased due to exponential damping and regularization. Consequently, the reconstructed signal captures the overall oscillation envelope but underestimates amplitude and phase fidelity. This is reflected in the spatiotemporal dynamics (Fig.~\ref{fig2}(f)), where periodic motion is qualitatively preserved but fine structure is degraded.

\subsection{Quantum-Coherent Dynamics}
Figure~\ref{fig3} highlights the role of phase coherence and its modification under imaginary-time evolution in the minimal nontrivial quantum system. In real time (Fig.~\ref{fig3}(a)), the two-level system exhibits coherent Rabi oscillations arising from relative phase evolution between energy eigenstates, encoding intrinsically dynamical information. By contrast, imaginary-time evolution (Fig.~\ref{fig3}(b)) progressively suppresses this phase structure and drives the system toward the ground state, reflecting its contractive and filtering character.
Despite this filtering, the simplicity of the spectrum makes the system highly amenable to inverse reconstruction. As shown in Fig.~\ref{fig3}(c), regularized inversion accurately recovers the dominant oscillatory dynamics over short and intermediate timescales, capturing both the oscillation frequency and amplitude with high fidelity. Deviations remain small and systematic, primarily reflecting the enhanced sensitivity of phase information to exponential spectral damping rather than a breakdown of the reconstruction itself.

Figure~\ref{fig4} illustrates the interplay between spectral near-degeneracy, imaginary-time filtering, and the stability of inverse reconstruction in a finite potential well. The bound-state spectrum (Fig.~\ref{fig4}(a)) contains closely spaced low-energy levels, giving rise to coherent tunneling dynamics. In real time (Fig.~\ref{fig4}(b)), an initially localized wave packet exhibits oscillatory motion driven by interference between nearly degenerate eigenstates, with a tunneling frequency set by the small energy splitting.
Under imaginary-time evolution (Fig.~\ref{fig4}(c)), this coherent behavior is progressively suppressed as the state becomes increasingly dominated by low-energy components. In the spectral representation (Fig.~\ref{fig4}(d)), all modes decay as $\exp(-E_n \tau/\hbar)$; however, the small energy gaps imply that low-lying states remain comparatively close in weight over a finite range of $\tau$, preserving the essential structure of the low-energy subspace before stronger compression sets in at larger imaginary times.
The inverse reconstruction (Fig.~\ref{fig4}(e)) demonstrates that this low-energy structure can be robustly recovered within a well-defined regime. The dominant spectral coefficients are reconstructed with good accuracy, allowing the essential features of the dynamics to be retained. Deviations increase gradually for higher-energy components and longer times, reflecting the growing conditioning challenge rather than a qualitative breakdown of the inversion. Accordingly, the reconstructed dynamics (Fig.~\ref{fig4}(f)) reproduces the characteristic tunneling timescale and captures the main oscillatory behavior, while finer details in amplitude and phase become progressively smoothed at larger $\tau$.

\subsection{Non-Hermitian Systems}
Figure~\ref{fig5} extends the analysis to non-Hermitian systems, where spectral non-normality introduces a qualitatively new layer of structure beyond exponential damping. The complex potential (Fig.~\ref{fig5}(a)) generates a non-Hermitian Hamiltonian with mixed real and imaginary eigenvalues, leading to gain–loss dynamics and non-orthogonal eigenstates. In real time (Fig.~\ref{fig5}(b)), the wave packet exhibits asymmetric evolution and amplitude modulation, reflecting the interplay between coherent propagation and dissipative or amplifying channels.
The corresponding contractive evolution (Fig.~\ref{fig5}(c)) suppresses spatial spreading and drives the system toward a localized profile. However, unlike the Hermitian case, this contraction is shaped not only by energy ordering but also by the non-orthogonality of the eigenmodes. In the spectral representation (Fig.~\ref{fig5}(d)), modal amplitudes undergo a non-uniform redistribution, resulting from the combined effects of damping and non-normal amplification, which encode the underlying spectral geometry.
The inverse reconstruction (Fig.~\ref{fig5}(e)) demonstrates that, despite this increased complexity, the global structure of the signal remains accessible. Regularization recovers the dominant features of the spectrum and yields a qualitatively accurate reconstruction of the dynamics. Differences emerge across a broader range of modes compared to the Hermitian case, particularly in intermediate components, reflecting the enhanced sensitivity associated with non-orthogonal eigenstates. Accordingly, the reconstructed signal captures the main dynamical patterns, while phase shifts and amplitude modulations reflect the intricate weighting of non-orthogonal contributions. This behavior is also evident in the spatiotemporal comparison (Fig.~\ref{fig5}(f)), where the reconstructed evolution retains the overall structure of the dynamics while exhibiting systematic deviations.

We do not present explicit reconstruction results for Lindblad master equations or relativistic Dirac dynamics, as these systems lie beyond the regime where inverse reconstruction remains quantitatively controlled. In open quantum systems governed by Lindblad generators, the evolution involves irreversible information transfer to the environment, resulting in genuine loss of coherence rather than purely spectral suppression. As a consequence, the forward map is no longer an information-compressive transformation, and even regularized inversion becomes limited.
A distinct obstruction arises in relativistic Dirac systems, whose spectrum is intrinsically indefinite, containing both positive- and negative-energy branches. Under imaginary-time evolution, these components behave qualitatively differently, with negative-energy modes undergoing unbounded amplification. Practical treatments therefore require projection onto the positive-energy subspace, which introduces additional information suppression and suppresses interference between energy sectors. The resulting inverse mapping is thus not only numerically unstable but restricted to a reduced dynamical subspace.
These considerations identify Lindblad and Dirac systems as limiting regimes in which reconstruction is constrained by intrinsic properties of the dynamics. Unlike the Hermitian and weakly non-Hermitian cases, where recoverability is governed by spectral conditioning and bandwidth, here the limitations originate from fundamental features of the generators—environment-induced irreversibility and spectral indefiniteness—placing strict bounds on the recoverable dynamical information.

\section{Conclusion}
A central result of this work is the identification of a 
recoverability structure governed by a logarithmic spectral cutoff 
and an exponential instability scaling.
Rather than viewing analytic continuation as merely ill-posed, 
we show that it exhibits a sharp, scale-dependent transition between 
recoverable and irrecoverable modes, characterized by
$E_c(\tau) \sim \frac{1}{\tau} \log \frac{1}{\epsilon}$.
This establishes a quantitative framework for analytic continuation, 
revealing it as a form of spectral compression with intrinsic limits 
analogous to bandwidth constraints in signal processing.
We have developed a unified spectral-semigroup framework that establishes an explicit analytic connection between real-time and imaginary-time quantum dynamics. 
Within this formulation, imaginary-time evolution emerges as a fractional spectral filter generated by a square-root operator, transforming unitary, phasecoherent dynamics into a contractive flow.
This perspective shows that the two descriptions are not merely related by Wick rotation, but are connected through a nonlocal operator acting on the temporal domain.
A central result is the identification of a scale-dependent asymmetry between forward evolution and inverse reconstruction. While imaginary-time dynamics defines a well-posed spectral compression that suppresses high-frequency components, the inverse mapping is generically ill-conditioned due to the amplification of these modes. Nevertheless, reconstruction remains feasible in a controlled spectral regime: across a broad class of systems—including continuous and discrete spectra, few-level coherent models, and non-Hermitian dynamics—we find that partial recovery remains feasible within a well-defined spectral bandwidth. The extent of this recoverability is controlled by spectral structure, with limitations arising from near-degeneracy, non-normality, and conditioning.
Beyond these regimes, we show that the limits of reconstruction can become structural. In open quantum systems and relativistic Dirac dynamics, mechanisms such as progressive loss of recoverable high-frequency information suppression and spectral indefiniteness impose intrinsic constraints that cannot be overcome by stabilization alone, restricting the accessible dynamical information.
Taken together, these results establish a unified picture in which analytic continuation acts as a spectral filtering process with well-defined and quantifiable limits on invertibility. Rather than a binary distinction between reversibility and irreversibility, the framework reveals a continuous transition governed by spectral structure, clarifying the fundamental limits of reconstruction in quantum dynamics.
The logarithmic cutoff and exponential scaling define a structure that applies across continuous, discrete, and non-Hermitian spectra.

\begin{acknowledgments}
This work was supported by the Adolf Martens Fellowship (Grant n. BAM-AMF-2025-1).
\end{acknowledgments}

\appendix
\section{Mathematical Subtleties and Scope of Validity}
While the spectral–semigroup formulation introduced above provides a unified description of analytic continuation, several technical aspects require clarification to ensure mathematical rigor. In the following, we refine key constructions and delimit their domain of validity.
The time-domain kernel representation introduced in Eqs.~(20)–(21),
\begin{equation}
	K(\tau,t') = \frac{1}{2\pi} \int_{-\infty}^{\infty} e^{-\omega \tau} e^{i\omega t'} \, d\omega,
\end{equation}
is formally suggestive but not well-defined as a Lebesgue integral, due to its divergence at large frequencies.
More precisely, this expression must be understood in a distributional sense, rather than as a pointwise-defined function. A mathematically controlled formulation requires specifying an appropriate test function space, such as the Schwartz space $\mathcal{S}(\mathbb{R})$, on which the kernel acts as a continuous linear functional:
\begin{equation}
	(K\psi)(\tau) = \langle K(\tau,\cdot), \psi(\cdot) \rangle, 
	\quad \psi \in \mathcal{S}(\mathbb{R}).
\end{equation}

Within this framework, the kernel should be interpreted as a tempered distribution obtained via analytic continuation in the complex frequency plane, and its meaning is rigorously defined only after suitable regularization (e.g., exponential damping or contour deformation).
Alternatively, a fully well-posed construction can be obtained by replacing the Fourier representation with a Laplace-transform-based formulation, or more generally within frameworks such as Fourier–Bros–Iagolnitzer (FBI) transforms, which provide analytic continuation in phase space.
We emphasize that throughout this work, all kernel expressions are to be interpreted as formal representations justified a posteriori by the spectral definition of the semigroup, which provides the primary mathematically rigorous foundation.
The square-root spectral deformation,
\begin{equation}
	\sqrt{A} = \int_{\sigma(A)} \sigma(\omega)\, dE_{\omega},
\end{equation}
relies on the functional calculus of the generator \( A = i\partial_t \).
To ensure mathematical consistency, this construction requires additional structural assumptions:
When \( A \) is self-adjoint, the definition follows directly from the spectral theorem, provided a consistent branch of the square root is chosen. The branch cut must be specified explicitly (e.g., along the negative real axis), and the requirement
\begin{equation}
	\Re \sigma(\omega) \ge 0
\end{equation}
ensures that the resulting operator generates a contractive semigroup.
In more general settings, the construction is rigorously justified when \( A \) is a sectorial operator with sector angle strictly less than \( \pi \). In this case, fractional powers \( A^\alpha \) (including \( \alpha = 1/2 \)) are well-defined via holomorphic functional calculus, for example through the Balakrishnan representation:
\begin{equation}
	A^\alpha x = \frac{\sin(\pi \alpha)}{\pi}
	\int_0^\infty \lambda^{\alpha-1}(A+\lambda I)^{-1} A x \, d\lambda.
\end{equation}

For non-normal generators, the lack of a projection-valued spectral measure implies that the above representation is no longer literal. Instead, one must rely on:
holomorphic functional calculus, and/or
semigroup-based definitions
In this regime, spectral representations should be viewed as effective or formal, and the resulting dynamics may depend sensitively on the choice of contour and resolvent bounds. In particular, the pseudospectrum may play a dominant role.
Accordingly, the results of this work are mathematically rigorous for self-adjoint or sectorial generators, and should be interpreted more cautiously for general non-normal operators.

We have identified the square-root deformation \( \alpha = 1/2 \) as a distinguished transformation interpolating between unitary (\( \alpha = 1 \)) and diffusive (\( \alpha = 0 \)) regimes.
More precisely, within the family of fractional generators \( A^\alpha \) for \( 0 < \alpha \le 1 \), the square-root case can be characterized as an intermediate scaling regime in which: high-frequency modes are suppressed sub-exponentially in the spectral parameter (e.g., \( e^{-\tau \sqrt{\omega}} \)), while low-frequency modes retain nontrivial dynamical weight.

However, the interpretation of \( \alpha = 1/2 \) as a “minimal deformation” should be understood in a structural rather than variational sense. The present work does not establish a formal optimality principle (e.g., minimization of a functional or entropy measure).
A more rigorous formulation would require identifying an extremal property, such as:
minimization of spectral distortion under contractivity constraints,
or an optimization problem balancing phase preservation and dissipation.
We leave such a variational characterization of fractional generators as an open problem for future work.

The present framework complements, rather than replaces, existing approaches to inverse analytic continuation. In particular:
Maximum Entropy (MaxEnt) and Bayesian methods incorporate prior information to stabilize inversion;  
Kernel ridge regression and Tikhonov schemes correspond to quadratic regularization in functional spaces;  
Neural-network-based approaches (e.g., deep learning reconstructions) attempt data-driven inversion of Laplace-type transforms.
In contrast, the approach developed here focuses on the operator-theoretic and spectral structure underlying all such methods, identifying an intrinsic recoverability bound:
\begin{equation}
	E_c(\tau,\epsilon) \sim \frac{1}{\tau} \log \frac{1}{\epsilon}.
\end{equation}

From this perspective, regularization schemes can be understood as practical implementations of a more fundamental constraint: the exponential decay of singular values of the forward operator and the resulting bandwidth-limited invertibility.
A detailed quantitative comparison with optimized reconstruction techniques lies beyond the scope of this work, but the spectral bounds derived here provide a universal baseline against which such methods may be evaluated.

\section{Fractional Generators and Functional Calculus}
The square-root operator introduced above can be understood as a fractional power of the generator within the framework of sectorial functional calculus. For a suitable sectorial operator $A$, its fractional powers $A^\alpha$ ($0<\alpha<1$) are well defined and generate strongly continuous semigroups.

A standard representation is provided by the Balakrishnan formula,
\begin{equation}
A^\alpha x = \frac{\sin(\pi \alpha)}{\pi} \int_0^\infty \lambda^{\alpha-1} (A+\lambda I)^{-1} A x \, d\lambda,
\end{equation}
which defines the operator in terms of the resolvent of $A$. In this sense, the square-root operator $\sqrt{A}$ corresponds to the case $\alpha = 1/2$.
Equivalently, the associated fractional semigroup can be represented through subordination. The evolution operator $e^{-\tau A^{1/2}}$ admits the integral representation
\begin{equation}
e^{-\tau A^{1/2}} = \frac{1}{2\sqrt{\pi}} \int_0^\infty \frac{\tau}{s^{3/2}} \exp\!\left(-\frac{\tau^2}{4s}\right) e^{-sA} \, ds,
\end{equation}
which expresses the square-root semigroup as a continuous superposition of exponential semigroups.

This representation is a special case of Lévy subordination, where the effective evolution time becomes a random variable distributed according to a stable law. From this perspective, the fractional dynamics arises from a stochastic time-change of an underlying Markovian semigroup.
These constructions place the square-root deformation within the standard theory of fractional semigroups, providing a rigorous basis for its role as an intermediate transformation between unitary and contractive dynamics.
This shows that the fractional deformation is not an ad hoc construction, but a canonical consequence of the functional calculus of sectorial operators.

\section{Spectral Filtering for the Imaginary-Time Evolution Operator}
The imaginary-time evolution operator admits a precise interpretation as a spectral filter. In the spectral representation of the generator $\mathcal{G}$, the evolution acts as
\begin{equation}
\hat{\psi}(\lambda) \;\mapsto\; e^{-\tau \lambda}\,\hat{\psi}(\lambda),
\end{equation}
so that the transfer function is given by $T_\tau(\lambda) = e^{-\tau \lambda}$.
This defines a monotone attenuation of modes with increasing spectral parameter $\lambda$. While the filter does not exhibit a sharp cutoff, it introduces an effective cutoff scale determined by the condition $e^{-\tau \lambda_c} \sim \epsilon$, yielding
\begin{equation}
\lambda_c(\tau,\epsilon) \sim \frac{1}{\tau}\log\frac{1}{\epsilon}.
\end{equation}
Modes with $\lambda \gg \lambda_c$ are exponentially suppressed and effectively unrecoverable in the presence of noise level $\epsilon$, while modes with $\lambda \lesssim \lambda_c$ are retained.
The attenuation rate is exponential in $\lambda$, $T_\tau(\lambda) \sim e^{-\tau \lambda}$
, which distinguishes this filter from algebraic or sharp cutoff filters and underlies the severe ill-conditioning of the inverse problem.

From a functional-analytic perspective, the operator $e^{-\tau \mathcal{G}}$ acts as a smoothing operator. In particular, it improves regularity by suppressing high-frequency components, mapping functions into smoother function classes. In typical settings, this corresponds to a gain in Sobolev regularity:
\begin{equation}
\| e^{-\tau \mathcal{G}} \psi \|_{H^s} \le C(\tau)\, \|\psi\|_{L^2},
\end{equation}
for $s>0$, reflecting the decay of high-frequency modes.
This smoothing property provides an alternative characterization of imaginary-time evolution as a low-pass filter: rather than sharply truncating the spectrum, it continuously redistributes spectral weight toward low-frequency components, with a strength controlled by $\tau$. In this sense, imaginary-time evolution acts as an exponentially decaying low-pass filter with a continuously tunable cutoff scale.

\section{Hilbert Space Setting}
Throughout this work, we consider dynamical evolution on a Hilbert space $\mathcal{H}$. The generator $\mathcal{G}$ is assumed to be a densely defined, closed linear operator on $\mathcal{H}$ that generates a strongly continuous ($C_0$) semigroup $\{K(\tau)\}_{\tau\ge 0}$.

In general, $G$ may be unbounded, and its domain $\mathcal{D}(\mathcal{G})\subset \mathcal{H}$ is specified as part of the operator. The evolution operator 
\begin{equation}
K(\tau) = e^{-\tau \mathcal{G}}
\end{equation}
is therefore understood in the sense of semigroup theory rather than as a naive operator exponential.

Depending on the system, the spectrum $\sigma(\mathcal{G})$ may be discrete, continuous, or mixed. In the self-adjoint or normal case, functional calculus is defined via the spectral theorem, with
\begin{equation}
\mathcal{G} = \int_{\sigma(\mathcal{G})} \lambda\, dE_\lambda ,
\end{equation}
where $E_\lambda$ is a projection-valued measure.
For continuous spectra, spectral integrals are understood in the sense of generalized eigenfunctions or direct integral decompositions, while in the discrete case they reduce to sums over eigenvalues. Mixed spectra combine both components.
Unless otherwise stated, expressions involving spectral integrals should be interpreted in this generalized sense.

Many physically relevant generators (e.g., Hamiltonians, differential operators) are unbounded. All operator expressions are therefore understood on appropriate domains, and equalities involving $\mathcal{G}$ or its functions are interpreted as holding on a dense invariant subspace.
To simplify notation across different systems, we denote by $\lambda$ a generic spectral parameter, which may represent energy, frequency, or other eigenvalues depending on the context. The associated spectral representation is written formally as
\begin{equation}
\psi = \int \hat{\psi}(\lambda)\, dE_\lambda,
\end{equation}
with the understanding that this notation encompasses both discrete sums and continuous integrals.



\begin{thebibliography}{99}
	
	\bibitem{1}
	E. C. G. Sudarshan and Charles B. Chiu,
	\textit{Analytic continuation of quantum systems and their temporal evolution},
	Phys. Rev. D, 47, 0602 (1993).
	
	\bibitem{2}
	Sarah Day, Jean-Philippe Lessard, and Konstantin Mischaikow,
	\textit{Validated Continuation for Equilibria of PDEs},
	SIAM J. Numer. Anal., 45, 1398-1424 (2007).
	
	\bibitem{3}
	N. Tanaka, Y. Suzuki, K. Varga, and R. G. Lovas,
	\textit{Unbound states by analytic continuation in the coupling constant},
	Phys. Rev. C, 59, 1391 (1999).
	
	\bibitem{4}
	O. Goulko, A. S. Mishchenko, L. Pollet, N. Prokof'ev, and Boris Svistunov,
	\textit{Numerical analytic continuation: Answers to well-posed questions},
	Phys. Rev. B, 95, 014102 (2017).
	
	\bibitem{5}
	P. Zhu, R. Wang, K. Sivagurunathan, S. Sfarra, F. Sarasini, C. Ibarra-Castanedo, X. Maldague, H. Zhang, and A. Mandelis,
	\textit{Frequency multiplexed photothermal correlation tomography for non-destructive evaluation of manufactured materials},
	Int. J. Extrem. Manuf., 7, 035601 (2025).
	
	\bibitem{6}
	P. Zhu, X. Maldague,
	\textit{THz-PINNs: Time-Domain Forward Modeling of Terahertz Spectroscopy With Physics-Informed Neural Networks},
	IEEE Trans. Terahertz Sci. Technol., 2026. DOI: 10.1109/TTHZ.2026.3674212
	
	\bibitem{7}
	P. Zhu, H. Zhang, S. Sfarra, F. Sarasini, C. Ibarra-Castanedo, X. Maldague, A. Mandelis,
	\textit{Thermal diffusivity measurement based on evaporative cryocooling excitation: Theory and experiments},
	arXiv:2509.04263 (2025).
	
	\bibitem{8}
	M. Fleischhauer, A. Imamoglu, and J. P. Marangos,
	\textit{Electromagnetically induced transparency: Optics in coherent media},
	Rev. Mod. Phys., 77, 633 (2005).
	
	\bibitem{9}
	F. D’Andrea, M. A. Kurkov, and F. Lizzi,
	\textit{Wick rotation and fermion doubling in noncommutative geometry},
	Phys. Rev. D, 94, 025030 (2016).
	
	\bibitem{10}
	F. Kuipers,
	\textit{Analytic continuation of stochastic mechanics},
	J. Math. Phys., 63, 042301 (2022).
	
	\bibitem{11}
	P. Zhu, J. Lecompagnon, P. D. Hirsch, M. Ziegler,
	\textit{Generalized Virtual-Wave Theory for Photothermal Coherence Tomography under Arbitrary Excitation Toward Non-Contact Industrial Inspection of Composite Materials},
	arXiv:2605.03747, 2026. https://doi.org/10.48550/arXiv.2605.03747 
	
	\bibitem{12}
	P. Zhu, J. Lecompagnon, M. Ziegler, C. Ibarra-Castanedo, X. Maldague,
	\textit{Generalized virtual wave reconstruction for vibrothermography: Overcoming the wavefront-free behavior and quantification challenges in the diffusion-wave field},
	arXiv:2603.23765, 2026. https://doi.org/10.48550/arXiv.2603.23765
	
	\bibitem{13}
	M. G. Cowling,
	\textit{Harmonic Analysis on Semigroups},
	Ann. Math., 117, 267-283 (1983).
	
	\bibitem{14}
	Barak Gabai and Xi Yin,
	\textit{Exact quantization and analytic continuation},
	J. High Energy Phys., 2023, 82 (2023).
	
	\bibitem{15}
	A. W. Sandvik,
	\textit{Stochastic method for analytic continuation of quantum Monte Carlo data},
	Phys. Rev. B, 57, 10287 (1998).
	
	\bibitem{16}
	D. Harlow, J. Maltz, and E. Witten,
	\textit{Analytic continuation of Liouville theory},
	J. High Energy Phys., 2011, 71 (2011).
	
	\bibitem{17}
	V. V. Kryzhniy,
	\textit{Numerical inversion of the Laplace transform: analysis via regularized analytic continuation},
	Inverse Probl., 22, 579 (2006).
	
	\bibitem{18}
	J. V. Herod and R. W. McKelvey,
	\textit{A Hille-Yosida theory for evolutions},
	Isr. J. Math., 36, 13-40 (1980).
	
	\bibitem{19}
	K. Symanzik,
	\textit{Euclidean Quantum Field Theory. I. Equations for a Scalar Model},
	J. Math. Phys., 7, 510-525 (1966).
	
	\bibitem{20}
	T. Lang, K. Liegener, and T. Thiemann,
	\textit{Hamiltonian renormalisation I: derivation from Osterwalder–Schrader reconstruction},
	Class. Quantum Grav., 35, 245011 (2018).
	
	\bibitem{21}
	N. Dungey,
	\textit{Subordinated discrete semigroups of operators},
	Trans. Amer. Math. Soc., 363, 2011 (2011).
	
	\bibitem{22}
	E. J. Hurtado,
	\textit{Non-local Diffusion Equations Involving the Fractional p(.)-Laplacian},
	J. Dyn. Differ. Equ., 32, 557-587 (2020).
	
	\bibitem{23}
	R. R. Coifman and G. Weiss,
	\textit{Extensions of Hardy spaces and their use in analysis},
	Bull. Amer. Math. Soc., 83, 569-645 (1977).
	
	\bibitem{24}
	A. Dienstfrey and L. Greengard,
	\textit{Analytic continuation, singular-value expansions, and Kramers-Kronig analysis},
	Inverse Probl., 17, 1307 (2001).
	
\end{thebibliography}

\end{document}